\documentclass[preprint,prd,aps,nofootinbib]{revtex4}
\usepackage{graphicx}
\usepackage{diagbox}
\usepackage{float}
\usepackage{mathrsfs}
\usepackage{color}
\usepackage{slashed}
\begin{document}
\title{Large Relativistic Corrections to Nonrelativistic $M1$ Transitions in Heavy Quarkonium}
\author{Su-Yan Pei$^{1,2,3}$,
Wei Li$^{4}$,
Wen-Yuan Ke$^{1,2,3}$,
Yi-Yi Rui$^{1,2,3}$,
Qiang Li$^5$, and
Guo-Li Wang$^{1,2,3}$\footnote{corresponding author}}

\affiliation{$^1$ Department of Physics, Hebei University, Baoding 071002, China
\nonumber\\
$^{2}$ Hebei Key Laboratory of High-precision Computation and Application of Quantum Field Theory, Baoding 071002, China
\nonumber\\
$^{3}$ Hebei Research Center of the Basic Discipline for Computational Physics, Baoding 071002, China
\nonumber\\
$^{4}$ College of Science, Hebei Agricultural University, Baoding 071001, China
\nonumber\\
$^5$ School of Physical Science and Technology, Northwestern Polytechnical University, Xi'an 710072, China}
\begin{abstract}
As double heavy quarkonia, charmonium and bottomonium are generally considered to have small relativistic corrections and can be treated using nonrelativistic models. However, this is not always the case.
In this paper, we employ the relativistic Bethe-Salpeter (BS) equation method to calculate the electromagnetic (EM) radiative decays of heavy quarkonium where the $M1$ transition provides the leading-order contribution. Compared to nonrelativistic method which only computes $M1$ transition, our calculations include $M1+E2+M3+E4$ transitions, where the higher-order multipoles, $E2$, $M3$, and $E4$, account for relativistic corrections. The study finds that relativistic effects are large in such transitions even for bottomonium. For instance: the relativistic corrections in the decays $\psi(nS)\rightarrow\gamma\eta_c(mS)$ ($n\geq m$) range from $68.1\%$ to $83.2\%$, while those for $\Upsilon(nS)\rightarrow\gamma\eta_b(mS)$ range between $65.9\%$ and $75.2\%$.
\end{abstract}
\maketitle
\section{Introduction}
Both charmonium and bottomonium are composed of double-heavy quarks. Given their heavy masses, charm quarks move slowly within charmonium systems, with their squared velocity conventionally estimated at $\langle v^2\rangle\approx 0.3$. Bottom quarks, being heavier, exhibit even slower motion in bottomonia, characterized by $\langle v^2\rangle\approx 0.1$. Due to such suppressed quark kinematics, charmonia and bottomonia are widely regarded to experience sufficiently small relativistic corrections to be accessible through non-relativistic theoretical frameworks.

However, this conclusion does not universally hold. First, this viewpoint is established based on research on the ground states, whereas for excited states, particularly for charmonium, significant deviations emerge. Our prior work reports $\langle v^2\rangle_{J/\psi}=0.26$ \cite{wgl}, consistent with the conventional 0.3 estimate, yet $\langle v^2\rangle_{\psi(4S)}=0.48$ and $\langle v^2\rangle_{\chi_{c2}(4P)}=0.51$ \cite{wgl} substantially exceeding 0.3, indicating non-negligible relativistic corrections in highly excited states. For instance, in the semileptonic weak decays $B_c\to\psi(nS)\ell\nu$ ($n=1,2,3$), relativistic effects contribute $19\%$, $37\%$, and $52\%$ respectively \cite{geng}. Second, the magnitude of relativistic corrections also depends on the types of interparticle interactions involved. This is exemplified in OZI-allowed strong decays: the process $B^*_2(5747)^0\to B^+\pi^-$ exhibits $13\%$ relativistic effects \cite{liu}, whereas $D^*_{s0}(2317)\to D^+_s\pi^0$ shows $64\%$ \cite{han}. Here, $B^*_2(5747)^0$ denotes the $2^+$ $B^*_2(1P)^0$ meson while $D^*_{s0}(2317)$ is the $0^+$ $D^*_{s0}(1P)$ meson.
Since all decays final states are $0^-$ pseudoscalars, the wave function overlap integral between the $2^+$ and $0^-$ mesons differs significantly from that between the $0^+$ and $0^-$ mesons. However, given this disparity, it remains challenging to establish general criteria for which processes require substantial relativistic corrections.

Another example, which also motivated us to initiate this study, concerns the radiative transitions of $\psi_2(1^3D_2)\to\chi_{c1}(1^3P_1)\gamma$ and $\psi_2(1^3D_2)\to\eta_{c}(1^1S_0)\gamma$. The former process is primarily an $E1$ transition at leading nonrelativistic order, while the latter is an $M1$ transition at leading order. We obtained that relativistic corrections account for $20\%$ in the former case, but reach as high as $68\%$ in the latter \cite{Li:2022qhg}. In contrast to nonrelativistic models that only compute leading-order $E1$ or $M1$ transitions, for decay $\psi_2(1^3D_2)\to\chi_{c1}(1^3P_1)\gamma$, we evaluate $E1+M2+E3$ transitions, while for $\psi_2(1^3D_2)\to\eta_{c}(1^1S_0)\gamma$, we calculate $M1+E2+M3$ contributions.
Our results demonstrate that in process $\psi_2(1^3D_2)\to\chi_{c1}(1^3P_1)\gamma$, the relativistic corrections ($M2+E3$) are significantly smaller than the nonrelativistic leading-order ($E1$) contribution, while in $\psi_2(1^3D_2)\to\eta_{c}(1^1S_0)\gamma$, the relativistic corrections ($E2+M3$) are comparable to or exceed the nonrelativistic leading-order ($M1$) contribution \cite{Li:2022qhg}.
Does this imply that processes with $M1$ transition as the leading-order universally exhibit large relativistic corrections?
Motivated by this fundamental question, we systematically investigate the following decays where $M1$ transition is the leading-order contribution, $\psi(nS,1D)\to\eta_c(mS)\gamma$, $\Upsilon(nS,1D)\to\eta_b(mS)\gamma$  ($n\geq m$), $\eta_c(nS)\to\psi(mS,1D)\gamma$, and $\eta_b(nS)\to\Upsilon(mS,1D)\gamma$  ($n > m$). The focus has been on relativistic corrections, namely, contributions from high-order multipoles transitions.

Regarding relativistic corrections for radiative transitions, some attempts have been made in the literature. For example, Refs. \cite{Brambilla:2005zw,Li:2011ssa} point out that the process of $J/\psi\rightarrow\gamma\eta_c$ has a large uncertainty due to high-order corrections \cite{Brambilla:2005zw,Li:2011ssa}.
Theoretical calculation \cite{Godfrey:2001eb} has revealed enormous relativistic corrections in the electromagnetic decays of low-lying bottomonia: $\Upsilon(2S)\rightarrow\gamma\eta_b(1S)$ and $\Upsilon(3S)\rightarrow\gamma\eta_b(1S, 2S)$. This conclusion is confirmed by \cite{Ebert:2002pp} through calculations of the processes $\eta_b(nS) \to \gamma\Upsilon(mS)$ and $\Upsilon(nS) \to \gamma\eta_b(mS)$ ($n>m$). Several experiments have searched for higher-order multipole contributions \cite{BaBar:2011xka,Zhao:2016pwu,Ai:2016eys,BESIII:2017tsq,CLEO:2009inp}, and the CLEO experiment reported significant $M2$ contribution to decays $\chi_{(c1,c2)}\to \gamma J/\psi$ \cite{CLEO:2009inp}.
The BaBar collaboration pointed out that higher-order relativistic corrections play a substantial role in hindered $M1$ transitions such as $\Upsilon(2S,3S)\rightarrow\gamma\eta_{b}(1S)$ \cite{BaBar:2011xka}.

At present, there have been significant developments in the exploration of the properties of heavy quarkonium \cite{Eichten:1979ms,Bodwin:1994jh,Brambilla:2010cs,Chen:2016qju,Olsen:2017bmm,Brambilla:2019esw,Strickland:2024oat},
including studies of electromagnetic decays \cite{Colquhoun:2023zbc,Deng:2015bva,Pineda:2013lta,Dudek:2009kk,Becirevic:2014rda,Barnes:2005pb,Brambilla:2005zw,Gaiser:1985ix}.
However, numerous low-lying excited states remain undiscovered, such as $\eta_b(2S)$ \cite{Feldman:1997qc,Barbieri:1979be,Belle:2012fkf}, $\Upsilon(1D)$ \cite{Radford:2007vd,Segovia:2016xqb,Deng:2016ktl}, and other states like $\psi(3S)$, $\eta_c(2S)$, $\Upsilon(2S)$, $\Upsilon(3S)$, whose electromagnetic decay processes require thorough theoretical and experimental investigation \cite{PDG}. So this study will enhance our understanding of these pseudoscalar and vector charmonia and bottomonia.

To investigate relativistic effects and provide precise theoretical predictions, this paper employs the Bethe-Salpeter (BS) equation method. The BS equation is a relativistic dynamic equation describing two body bound state  \cite{Salpeter:1951sz}, and Salpeter equation is its instantaneous version \cite{Salpeter:1952ib}, {suitable for heavy mesons.
Unlike non-relativistic approaches such as Heavy Quark Effective Theory (HQET) or Non-Relativistic QCD (NRQCD), which employ expansion methods for handling relativistic corrections, expanding in powers of the inverse heavy quark mass $1/m_Q$ or the quark velocity $v$ (momentum $q = mv$), and typically include only the first or second-order relativistic corrections, thereby failing to guarantee computational accuracy, the BS equation and the Salpeter equation are integral equations. The effect of iterative integration is equivalent to expanding $q$ to infinite orders in the expansion method, thereby encompassing the complete relativistic corrections of the expansion approach. Although the BS equation method has advantages in relativistic calculations, similar to other existing theories, it also has limitations, such as it does not account for contributions from higher Fock states, multi-quark states, or coupled-channel effects.}

This paper will solve the complete Salpeter equation comprising four sub-equations. The resulting relativistic wave function incorporates distinct partial wave components, thereby accounting for $M1+E2+M3+E4$ transitions in the calculation of electromagnetic radiative decays between vector and pseudoscalar mesons.
This method has been widely applied in electromagnetic decay processes \cite{Wang:2012cp,Wang:2015yea,Li:2022qhg,Pei:2022cjy,Li:2023cpl,Du:2024gna,Pei:2024hzv} and is well established as a reliable approach; therefore, it will not be elaborated further here.

The organization of this paper is as follows. Sec. II presents our theoretical framework, including the formulation of hadronic transition matrix elements, relativistic wave functions incorporating different partial waves, and the multipole contributions to the transition amplitude arising from overlap integrals between partial waves in the initial and final state wave functions. Sec. III provides the results of the electromagnetic decays and the contributions from different multipole transitions. Finally, we give a summary in Sec. IV.
\section{The theoretical method}
In this section, we will present the calculation formula for the electromagnetic radiative transition amplitude, introduce the relativistic wave functions used, and discuss the different partial waves within the wave functions. The classification of multipole transition amplitudes within the amplitude will also be discussed in detail.
\subsection{Electromagnetic radiative transition matrix elements}
\begin{figure}[h]
\centering
\includegraphics{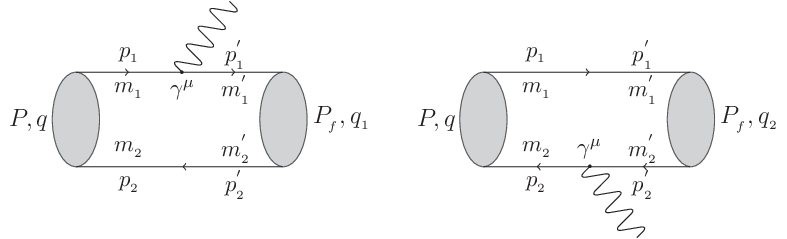}
\caption{Feynman diagrams for the EM transition}\label{feyn}
\end{figure}

A quarkonium, composed of a quark with momentum $p_1$ and an antiquark with momentum $p_2$, has a total momentum $P$ and an internal relative momentum $q$. These momenta satisfy the following relations:
$$p_1=\frac{1}{2}P+q,~p_2=\frac{1}{2}P-q.$$
When the quarkonium undergoes an electromagnetic radiation transition, the photon can be emitted by either the quark or the antiquark. Therefore, this process involves two Feynman diagrams, as shown in Figure \ref{feyn}.  The $S$-matrix element for this transition process can be written as:
\begin{eqnarray}
&&T=<P_{_{f}}\epsilon_{_f},k\epsilon_{_0}|S|P\epsilon>=(2\pi)^{4}\delta^{4}(P_{_{f}}+k-P)\epsilon_{_{0}\mu}\mathcal{M}^{\mu},
\end{eqnarray}
where $P$, $P_{f}$, and $k$ are the momenta of initial meson, the final meson, and the photon, respectively; $\epsilon_{_0}$, $\epsilon$, and $\epsilon_{_f}$ are the polarization vectors of the photon, the initial and the final quarkonium, respectively. $\mathcal{M}^{\mu}$ is the hadronic transition matrix element, which is a model-dependent quantity.

Within the framework of the BS equation approach, we express it in terms of the overlap integral of the relativistic wave functions of the initial and final mesons as follows \cite{chc}:
\begin{eqnarray}\label{trans}
\mathcal{M}^{\mu}=\int\frac{d^{3}q_{_{\bot}}}{(2\pi)^{3}}\left\{Tr\left[\frac{\slashed{P}}{M}\overline{\varphi}_{_{P_{_{f}}}}^{++}(q_{_{1\bot}})Q_{_{1}}\gamma^{\mu}\varphi_{_{P}}^{++}(q_{_{\bot}})
+\overline{\varphi}_{_{P_{_{f}}}}^{++}(q_{_{2\bot}})\frac{\slashed{P}}{M}\varphi_{_{P}}^{++}(q_{_{\bot}})Q_{_{2}}\gamma^{\mu}\right]\right\},
\end{eqnarray}
where $M$ is the mass of initial meson, $\varphi_{_{P}}^{++}$ and $\varphi_{_{P_{_{f}}}}^{++}$ are the positive energy wave functions of initial and final mesons, and $\overline{\varphi}_{_{P_{_{f}}}}^{++}=\gamma_0{({\varphi}_{_{P_{_{f}}}}^{++})}^{\dagger}\gamma_0$. We have neglected the contributions from negative energy wave functions and similar terms with minimal contributions. For charmonium, $Q_{_{1}}=\frac{2}{3}e$ and $Q_{_{2}}=-\frac{2}{3}e$, while for bottomonium, $Q_{_{1}}=-\frac{1}{3}e$ and $Q_{_{2}}=\frac{1}{3}e$, $e$ is the unit charge. Relative momentum $q$ can be divided into two parts, $q_{_{\parallel}}$ and $q_{_{\bot}}$,  where $q_{_{\parallel}}^{\mu}\equiv(P\cdot q/M^{2})P^{\mu}$, and $q_{_{\bot}}^{\mu}\equiv q^{\mu}-q_{_{\parallel}}^{\mu}$. In the center-of-mass frame of the meson, they become to $q_{0}$ and $\vec{q}$, respectively. The relative momenta of the final meson satisfy the following relations with that of the initial meson, $q_{_{1\bot}}=q_{_{\bot}}+\frac{1}{2}P_{f_{\bot}}$ and $q_{_{2\bot}}=q_{_{\bot}}-\frac{1}{2}P_{f_{\bot}}$.

{In the non-relativistic limit, the electromagnetic transition of $1^-\to 0^-\gamma$ involves only the $M1$ transition, and its transition amplitude is proportional to the overlap integral of the wave functions of the initial and final mesons \cite{em1,em}:
\begin{eqnarray}\label{M1}
\mathcal{M}_{M1}\propto\int{d^{3}q_{_{\bot}}}R_{0^-}(q_{_{\bot}})\cdot1\cdot R_{1^-}(q_{_{\bot}}),
\end{eqnarray}
where $R_{0^-}(q_{_{\bot}})$ and $R_{1^-}(q_{_{\bot}})$ are the normalized radial wave functions of $0^-$ and $1^-$ mesons, respectively.
However, in our relativistic approach, see Eq.(\ref{trans}), the transition amplitude includes not only the lowest-order $M1$ transition but also higher-order transitions such as $E2$, $M3$, and $E4$, which are expressed as,
\begin{eqnarray}\label{E2}
\mathcal{M}^{\alpha}_{E2}\propto\int{d^{3}q_{_{\bot}}}R_{0^-}(q_{_{\bot}})\cdot q_{_{\bot}}^{\alpha}\cdot R_{1^-}(q_{_{\bot}}),
\end{eqnarray}
\begin{eqnarray}\label{M3}
\mathcal{M}^{\alpha\beta}_{M3}\propto\int{d^{3}q_{_{\bot}}}R_{0^-}(q_{_{\bot}})\cdot q_{_{\bot}}^{\alpha}q_{_{\bot}}^{\beta}\cdot R_{1^-}(q_{_{\bot}}),
\end{eqnarray}
\begin{eqnarray}\label{E4}
\mathcal{M}^{\alpha\beta\gamma}_{E4}\propto\int{d^{3}q_{_{\bot}}}R_{0^-}(q_{_{\bot}})\cdot q_{_{\bot}}^{\alpha}q_{_{\bot}}^{\beta}q_{_{\bot}}^{\gamma}\cdot R_{1^-}(q_{_{\bot}}).
\end{eqnarray}
}

After performing the trace and integrating out the internal relative momenta, the hadronic transition matrix element in Eq. (\ref{trans}) is expressed in terms of form factors as follows:
\begin{eqnarray}\label{shape}
\mathcal{M}^{\mu}_{_{1^{--}\rightarrow0^{-+}}}=&&ia_1\varepsilon^{\mu\epsilon P P_{_{f}}},
\nonumber\\
\mathcal{M}^{\mu}_{_{0^{-+}\rightarrow1^{--}}}=&&ia_2\varepsilon^{\mu\epsilon_{_{f}} P P_{_{f}}},
\end{eqnarray}
where $i$ denotes the imaginary unit, $\varepsilon$ represents the Levi-Civita symbol, and $a_1$ and $a_2$ are the form factors. The subscript $1^{--} \to 0^{-+}$ represents the decay of a vector quarkonium into a pseudoscalar quarkonium, while $0^{-+} \to 1^{--}$ denotes the transition from a pseudoscalar quarkonium to a vector quarkonium. We have also used the following abbreviation $\varepsilon^{\mu\epsilon P P_{_{f}}}=\varepsilon^{\mu\nu\alpha\beta}\epsilon_{_\nu} P_{_\alpha} P_{_{f\beta}}$.

The decay width for the electromagnetic radiative process is
\begin{eqnarray}
\Gamma=\frac{|\vec{P_{_f}}|}{8\pi M^{2}}\frac{1}{2J+1}\sum{|\epsilon_{0\mu}\mathcal{M^{\mu}}|^{2}},
\end{eqnarray}
where $|\vec{P_{_f}}|=(M^2-M^2_{_f})/{2M}$ is the three-dimensional momentum of the final quarkonium, $M_{_f}$ is the mass of final meson, and $J$ is the total angular momentum of initial meson.
\subsection{The positive wave functions}
In this paper, we will continue to use the conventional notation $n^{2S+1}L_{J}$ (or abbreviated as $nL$), commonly used in the literature, to distinguish particles. Here, $n$ is the principal quantum number, $S$ is the spin, $L$ is the orbital angular momentum, and $J$ is the total angular momentum. However, we need to point out that this description is based on the non-relativistic framework of atomic physics. In this descriptive approach, the particle's wave function is constructed from a single wave, corresponding to $L=0, 1, 2...$ (i.e., $S$-wave, $P$-wave, $D$-wave...).

Yet, we note that particles possess definite $J^P$ quantum numbers (where $P$ is parity, defined as $P=(-1)^{L+1}$), which are intrinsic properties independent of whether a non-relativistic or relativistic method is used to describe the particle. Therefore, we choose to present the relativistic representation of the particle wave function based on $J^P$  \cite{Wang:2022cxy}, { not the $n^{2S+1}L_{J}$. Consequently, the wave functions obtained using this approach no longer consist of a single partial wave but inherently incorporate different partial wave components. That is to say, in our method, although the meson's wave function corresponds to a definite total angular momentum $J$, the orbital angular momentum $L$ is not a good quantum number. Consequently, the wave function does not correspond to a single, unique orbital angular momentum but rather encompasses multiple orbital angular momentum components. Due to this property of the wave function, the transition amplitudes derived from its application include a greater number of multipole amplitudes.}

For vector meson, whose $J^P=1^{-}$, the general form of the relativistic wave function can be constructed using the total momentum $P$, the internal relative momentum $q$, the polarization vector $\epsilon$, gamma matrix, and radial wave function, etc.
We combine these physical quantities to construct a general relativistic wave function representation $\varphi$ with quantum number $1^-$, consisting of a total of 16 terms. However, we do not intend to present the expression for $\varphi$. Instead, we show the positive-energy wave function $\varphi^{++}$. $\varphi^{++}$ is derived directly from $\varphi$ (see Reference \cite{Wang:2005qx} for details), and they share an identical structure.
Since we have chosen the instantaneous approximation, 8 terms containing $P\cdot q=P\cdot q_{_{\bot}}$ become zero. Then the positive energy wave function of the $1^{--}$ ($\psi$ or $\Upsilon$) meson \cite{Wang:2005qx} is written as,
\begin{eqnarray}\label{1-}
\varphi_{1^{--}}^{++}(q_{\bot})=&&q_{\bot}.\epsilon\left(A_{1}+\frac{\slashed{P}}{M}A_{2}+\frac{\slashed{q}_{\bot}}{M}A_{3}+\frac{\slashed{P}\slashed{q}_{\bot}}{M^{2}}A_{4}\right)
\nonumber\\
&&+M\slashed{\epsilon}\left(A_{5}+\frac{\slashed{P}}{M}A_{6}+\frac{\slashed{q}_{\bot}}{M}A_{7}+\frac{\slashed{P}\slashed{q}_{\bot}}{M^{2}}A_{8}\right),
\end{eqnarray}
$A_{i} (i=1\sim8)$ are related to the four independent radial wave functions $a_{3}$, $a_{4}$, $a_{5}$, and $a_{6}$ of $\varphi$, which are functions of $-{q}_{\bot}^{2}$. {In this case, Eq.(\ref{1-}) represents the most general form of the relativistic wave function (in the instantaneous approximation) for a $1^-$ meson, encompassing all possibilities. That is to say, all higher-order $-{q}_{\bot}^{2}$ contributions are incorporated into the radial wave functions. Therefore, this wave function effectively includes results equivalent to expanding $q_{\bot}$ to infinite orders in the expansion method.} By solving the Salpeter equation, we can obtain the numerical solution of the radial wave functions. And they have the following relationships,
$$f=\frac{1}{2}\left(a_3+\frac{m}{w}a_4\right),~~
A_{1}=\frac{q_{_\bot}^{2}}{Mm}f+\frac{M}{2m}\left(a_{5}-\frac{m}{w}a_{6}\right),~~
A_{5}=\frac{1}{2}\left(a_{5}-\frac{w}{m}a_{6}\right),~~$$
$$A_{2}=-\frac{M}{w}A_5,~
A_{3}=f-\frac{M^{2}}{2mw}a_{6},~
A_{4}=\frac{w}{m}f-\frac{M^{2}}{2mw}a_{5},~
A_{6}=-\frac{m}{w}A_5,~
A_{7}=0,~
A_{8}=A_2,$$
where $m$ and $w=\sqrt{m^2-q_{_\bot}^2}$ are the mass and energy of the quark (antiquark), respectively.

Next, we will explain that the wave function of the vector meson based on the $J^P$ assignment is not a pure $S$ wave, but rather a mixture of $S$, $P$, and $D$ waves. When we transform the coordinate system to spherical coordinates, the positive-energy wave function of $1^{--}$ in Eq. (\ref{1-}) becomes
\cite{Chang:2004im},
\begin{eqnarray}\label{p1-}
\varphi_{1^{--}}^{++}(q_{\bot})=&&M\sqrt{\frac{4\pi}{3}}C_{\lambda}\left(A_{1}+\frac{\slashed{P}}{M}A_{2}\right)+\left(\frac{1}{3}q_{_\bot}^{2}A_{\lambda}-\frac{2}{3}\sqrt{\frac{\pi}{5}}q_{_\bot}^{2}B_{\lambda}\right)\left(\frac{1}{M}A_{3}-\frac{\slashed{P}}{M^{2}}A_{4}\right)
\nonumber\\
&&+MA_{\lambda}\left(A_{5}+\frac{\slashed{P}}{M}A_{6}\right)-\frac{E_{\lambda}\slashed{P}}{M}A_{8},
\end{eqnarray}
where
$$A_{\lambda}=\epsilon^{+}\gamma^{-}+\epsilon^{-}\gamma^{+}-\epsilon^{\Delta}\gamma^{\Delta},~~C_{\lambda}=\frac{\vert\vec q\vert}{M}\left(\epsilon^{-}Y_{11}+\epsilon^{+}Y_{1-1}-\epsilon^{\Delta}Y_{10}\right),$$
$$~~B_{\lambda}=\sqrt{6}\epsilon^{+}\gamma^{+}Y_{2-2}-\sqrt{3}\left(\epsilon^{+}\gamma^{\Delta}+\epsilon^{\Delta}\gamma^{+}\right)Y_{2-1}+\left(2\epsilon^{\Delta}\gamma^{\Delta}+\epsilon^{-}\gamma^{+}+\epsilon^{+}\gamma^{-}\right)Y_{20}$$
$$-\sqrt{3}\left(\epsilon^{-}\gamma^{\Delta}+\epsilon^{\Delta}\gamma^{-}\right)Y_{21}+\sqrt{6}\epsilon^{-}\gamma^{-}Y_{22},$$
$$~~E_{\lambda}=2\sqrt{\frac{\pi}{3}}|\vec{q}|\left(\epsilon^{\Delta}\gamma^{\Delta}-\epsilon^{-}\gamma^{+}-\epsilon^{+}\gamma^{-}\right)\left(\gamma^{\Delta}Y_{10}-\gamma^{-}Y_{11}-\gamma^{+}Y_{1-1}\right),$$
and $\gamma^{+}\equiv-\frac{\gamma^{1}+i\gamma^{2}}{\sqrt{2}},~\gamma^{-}\equiv\frac{\gamma^{1}-i\gamma^{2}}{\sqrt{2}},$ $Y_{lm}$ are spherical harmonics. Therefore, the wave function $\varphi_{1^{--}}^{++}(q_{\bot})$ we present contains an $S$ wave component $$
\varphi^S_{1^{--}}=M\slashed{\epsilon}\left(A_{5}+\frac{\slashed{P}}{M}A_{6}\right)
+\frac{1}{3}q_{_\bot}^{2}\slashed{\epsilon}\left(\frac{1}{M}A_{3}-\frac{\slashed{P}}{M^{2}}A_{4}\right),
$$
a $P$ wave component $$\varphi^P_{1^{--}}=q_{\bot}.\epsilon\left(A_{1}+\frac{\slashed{P}}{M}A_{2}\right)
+\frac{\slashed{\epsilon}\slashed{P}\slashed{q}_{\bot}}{M}A_{8},$$
and a $D$ wave component $$\varphi^D_{1^{--}}=\left(\epsilon\cdot q_{_\bot}\slashed{q}_{\bot}-\frac{1}{3}q_{_\bot}^{2}\slashed{\epsilon}\right)
\left(\frac{1}{M}A_{3}-\frac{\slashed{P}}{M^{2}}A_{4}\right).$$
That is,
\begin{eqnarray}
\varphi_{1^{--}}^{++}=\varphi^S_{1^{--}}+\varphi^P_{1^{--}}+\varphi^D_{1^{--}}.
\end{eqnarray}
{Therefore, it is evident that for a vector meson with a definite total angular momentum $J=1$, only the $S$-wave ($L=0$) is present in the non-relativistic case, and the wave function is $M\slashed{\epsilon}\left(A_{5}+\frac{\slashed{P}}{M}A_{6}\right)$. In this scenario, there is a one-to-one correspondence between the total angular momentum and the orbital angular momentum. In contrast, its relativistic wave function includes not only the $S$-wave but also $P$-wave ($L=1$) and $D$-wave ($L=2$) components, meaning that a single total angular momentum corresponds to multiple orbital angular momenta. This leads us to obtain more multipole amplitudes in the transition amplitudes.}

{
Experiments have revealed that the meson $\psi(3770)$ can decay into lepton-antilepton pairs \cite{PDG}, which contradicts the assumption of a pure $D$-wave state, indicating that it must contain an $S$-wave component. Therefore, it is widely believed that $\psi(2S)$ and $\psi(3770)$ are $S-D$ wave mixed states. This phenomenon demonstrates, on one hand, that orbital angular momentum $L$ is indeed not a good quantum number, and on the other hand, it shows that the total angular momentum $J$ of a particle does not have a one-to-one correspondence with its orbital angular momentum $L$, but rather a one-to-two relationship. Our results for the $1^-$ state go even further, revealing a one-to-three relationship.}

The general form of the pseudoscalar meson is constructed from $P$, $q_{\bot}$, and gamma matrix. Because of the instantaneous approximation, the general form of the relativistic wave function for scalar mesons is reduced from 8 terms to 4 terms. The positive energy wave function of the $0^{-+}$ ($\eta_{_{c}}$ or $\eta_{_{b}}$) meson \cite{Kim:2003ny} can be written as:
\begin{eqnarray}\label{s0-}
\varphi_{0^{-+}}^{++}(q_{\bot})=\left(B_{1}+B_{2}\frac{\slashed{P}}{M}+B_{3}\frac{\slashed{q}_{\bot}}{M}+B_{4}\frac{\slashed{q}_{\bot}\slashed{P}}{M}\right)\gamma_{5},
\end{eqnarray}
where $B_i$ is the function of radial wave function $b_i~(i=1,2)$,
$$
B_{1}=\frac{M}{2}\left(b_{1}+\frac{m}{w}b_{2}\right),~~B_{2}=\frac{w}{m}B_{1},~~B_{3}=0,~~B_{4}=\frac{1}{m}B_{1}.
$$
And the numerical values of the independent radial wave functions $b_1$ and $b_2$ are obtained by solving the Salpeter equation \cite{Kim:2003ny}.

Eq. (\ref{s0-}) also indicates that the $0^{-+}$ meson wave function is a mixture of $S$ wave and $P$ wave. The $S$ wave component,
$$\varphi_{0^{-+}}^{S}=\left(B_{1}+B_{2}\frac{\slashed{P}}{M}\right)\gamma_{5},$$
and the $P$ wave component,
$$\varphi_{0^{-+}}^{P}(q_{\bot})=\left(B_{4}\frac{\slashed{q}_{\bot}\slashed{P}}{M}\right)\gamma_{5}$$
So we have
\begin{eqnarray}
\varphi_{0^{-+}}^{++}=\varphi^S_{0^{-+}}+\varphi^P_{0^{-+}}.
\end{eqnarray}
\subsection{The Electromagnetic Multipole Contributions}
According to the parity selection rule, magnetic transition $M_{L}$ satisfies $\pi_{i}\pi_{f}=(-1)^{L+1}=(-1)^{\pi_{i}}(-1)^{\pi_{f}}$, and electric  transition $E_{L}$ satisfies $\pi_{i}\pi_{f}=(-1)^{L}=(-1)^{\pi_{i}}(-1)^{\pi_{f}}$, where $\pi_{i}$ and $\pi_{f}$ represent the parity of the initial and final state mesons, respectively. We distinguish electromagnetic multipole transitions by selecting different partial wave in the wave function.

As can be seen from Eq. (\ref{trans}), the electromagnetic radiative transition amplitude is the overlap integral of the initial and final state wave functions. Given that our wave function contains distinct partial wave components, following the classification above, we can decompose the transition amplitude into a sum of electric and magnetic transitions of different multipole orders.
Taking the electromagnetic decay of vector to pseudoscalar as an example, the initial state wave function is $\varphi_{1^{--}}^{++} = \varphi^S_{1^{--}} + \varphi^P_{1^{--}} + \varphi^D_{1^{--}}$, denoted as $S + P + D$. The final state is $\varphi_{0^{-+}}^{++} = \varphi^S_{0^{-+}} + \varphi^P_{0^{-+}}$, denoted as $S' + P'$, where primes are used on final state partial waves to distinguish them from initial state components.

For pseudoscalar meson, only the $S$ wave exists in the non-relativistic limit, making the $P$ wave component a relativistic correction. For $S$ wave dominated vector charmonium $\psi(nS)$ ($n=1,2,3$), the non-relativistic case likewise contains only $S$ wave, with first order relativistic correction from $P$ wave and second order correction from $D$ wave.
Consequently, in the $^3S_1 \to{}^1S_0$ transition amplitude, the leading-order term is the non-relativistic magnetic dipole transition $M1 = S\times S'$, {which is just the with relativistic corrections emerging at second order as the electric quadrupole $E2 = S\times P' + P\times S'$, at third order as the magnetic octupole $M3 = P\times P' + D\times S'$, and at fourth order as the electric hexadecapole $E4 = D\times P'$.}

\begin{table}[htp]
\footnotesize
\begin{center}
\caption{The EM multipole contributions.} \label{incharm}
\begin{tabular}{|c|c|c|c|c|}
\hline\hline
\textbf{Process}           &\textbf{M1}     &\textbf{E2}       &\textbf{M3 }       &\textbf{E4}     \\
\hline
~~~~~$n^{3}S_{1}\rightarrow\gamma m^{1}S_{0}$~~~~~&$S\times S'$   &$S\times P'+P\times S'$&~~~~~~$P\times P'+D\times S'$~~~~~~&$D\times P'$       \\
$n^{3}D_{1}\rightarrow\gamma m^{1}S_{0}$    & ~~~~~$D\times S'$~~~~~&$D\times P'+P\times S'$     &$S\times S'+P\times P'$&~~~~~~$S\times P'$~~~~~~ \\
$n^{1}S_{0}\rightarrow\gamma m^{3}S_{1}$          &$S\times S'$&~~~~~~$S\times P'+P\times S'$~~~~~~&$S\times D'+P\times P'$     &$P\times D'$       \\
$n^{1}S_{0}\rightarrow\gamma m^{3}D_{1}$          &$S\times D'$     &$S\times P'+P\times D'$     &$S\times S'+P\times P'$     &$P\times S'$         \\
\hline\hline
\end{tabular}
\end{center}
\end{table}

These corresponding results are summarized in Table \ref{incharm}. From the table, it is evident that for both ${}^3S_1 \to {}^1S_0$ and ${}^1S_0 \to {}^3S_1$ decays, the transition amplitudes calculated by our method consist of $M1 + E2 + M3 + E4$ components. While unlike vector $\psi(nS)$ state, the $D$ wave in the $\psi(1D)$ wave function provides the zeroth-order non-relativistic contribution, with the first-order relativistic correction arising from the $P$ wave component, while the $S$ wave contributes as a second-order relativistic correction.

\section{Numerical Results and discussions}
In our calculations, there are model-dependent parameters, such as $m_b=5.2~\mathrm{GeV}$, $m_c=1.4~\mathrm{GeV}$, and so forth. We do not elaborate here on the details of solving the complete Salpeter equation; interested readers may refer to references \cite{Chang:2010kj,wglwu}. Regarding the meson masses, we match the experimental values by adjusting the free parameter when solving the equation. And we treat the particle $X(3940)$ as $\eta_c(3S)$, whose mass is $3942$ MeV. For particles without experimental values, we directly adopt the masses predicted by our model, e.g., $M_{_{\Upsilon(1D)}}=10129$ \rm{MeV}, and $\eta_b(3S)=10336$ \rm{MeV}.

\subsection{The EM decays of charmonium}
In Table \ref{charm}, we present the decay widths for radiative transition of charmonium. For comparative purposes, we also include the experimental data from Particle Data Group (PDG) \cite{PDG}, and theoretical predictions from the covariant Blankenbecler-Sugar (BSLT) equation \cite{Lahde:2002wj}, the relativistic quark model \cite{Ebert:2002pp}, the heavy quark effective theory \cite{Wang:2011rt}, the effective Lagrangian approach designed to incorporate contributions from intermediate meson loops \cite{Li:2007xr,Li:2011ssa}, and the Nonrelativistic QCD (NRQCD) \cite{QuarkoniumWorkingGroup:2004kpm}.

\begin{table}[htp]
\begin{center}
\caption{The EM decay widths (keV) of charmonium.} \label{charm}
\begin{tabular}{c c c c c c c c}
\hline\hline
~~~~~~~~~\textbf{Process}~~~~~~~~~&~~~\textbf{Ours}~~~&~~~\textbf{\cite{Lahde:2002wj}}~~~&~~~\textbf{\cite{Ebert:2002pp}}~~~ &~~~\textbf{\cite{Wang:2011rt}}~~~&~~~\textbf{\cite{Li:2007xr,Li:2011ssa}}~~~&~~~\textbf{\cite{QuarkoniumWorkingGroup:2004kpm}}~~~    &~~~\textbf{PDG \cite{PDG}}~~~\\
\hline
$\psi(1S)\rightarrow\gamma\eta_c(1S)$   &1.46    &1.25 &1.05  &      &$1.58\pm0.37$  &1.96  &$1.31\pm 0.15$ \\
\hline
$\psi(2S)\rightarrow\gamma\eta_c(1S)$   &1.20    &1.13 &0.95  &0.089 &$2.05^{+2.65}_{-1.75}$&0.926 &$1.05\pm 0.18$   \\
$\psi(2S)\rightarrow\gamma\eta_c(2S)$   &0.135   &0.03 &0.043 &   &$0.08\pm0.03$ &0.14  &$0.21\pm 0.15$   \\
\hline
$\psi(3770)\rightarrow\gamma\eta_c(1S)$ &1.54    &     &   &   &$17.14^{+22.93}_{-12.03}$&   &$<19$    \\
$\psi(3770)\rightarrow\gamma\eta_c(2S)$ &0.0233  &     &   &   &$1.82^{+1.95}_{-1.19}$&   &$<24$    \\
\hline
$\psi(4040)\rightarrow\gamma\eta_c(1S)$ &0.0122  &     &      &0.244 &      &      &                     \\
$\psi(4040)\rightarrow\gamma\eta_c(2S)$ &0.140   &     &      &0.021 &      &      &                     \\
$\psi(4040)\rightarrow\gamma\eta_c(3S)$ &0.864   &     &      &      &      &      &                     \\
\hline
$\eta_c(2S)\rightarrow\gamma\psi(1S)$   &0.630   &1.83 &1.53  &0.136 &      &0.538 &$<1.7$             \\
\hline
$\eta_c(3S)\rightarrow\gamma\psi(1S)$   &0.413   &     &      &0.430 &      &      &                     \\
$\eta_c(3S)\rightarrow\gamma\psi(2S)$   &0.112   &     &      &0.018 &      &      &                     \\
$\eta_c(3S)\rightarrow\gamma\psi(1D)$   &0.00316 &     &      &      &      &      &                     \\
\hline\hline
\end{tabular}
\end{center}
\end{table}

It can be seen from Table \ref{charm} that our calculated decay widths exhibit good agreement with the experimental data from PDG. In addition, the branching ratio $Br({\psi(1S)\rightarrow\gamma\eta_c(1S)})=1.58\%$ we calculated is close to the data $(1.98\pm0.09\pm0.30)\%$ given by CLEO \cite{CLEO:2008pln}. And our result $Br({\psi(2S)\rightarrow\gamma\eta_c(1S)})=4.10\times10^{-3}$, which exhibits excellent agreement with the data $(4.32\pm0.16\pm0.60)\times10^{-3}$ obtained by CLEO \cite{CLEO:2008pln}. The BESIII has searched for radiative decay channels $\psi(3770)\rightarrow\gamma\eta_c(1S)$ and $\psi(3770)\rightarrow\gamma\eta_c(2S)$, establishing upper limits for their branching ratios as $Br({\psi(3770)\rightarrow\gamma\eta_c(1S)})<6.8\times10^{-4}$ and $Br({\psi(3770)\rightarrow\gamma\eta_c(2S)})<2.0\times10^{-3}$ at $90\%$ confidence level \cite{BESIII:2014nsw}, which are consistent with our theoretical predictions $Br({\psi(3770)\rightarrow\gamma\eta_c(1S)})=5.66\times10^{-5}$ and $Br({\psi(3770)\rightarrow\gamma\eta_c(2S)})=8.57\times10^{-7}$.

In theoretical studies, our result for $\psi(1S)\rightarrow\gamma\eta_c(1S)$ is close to the predictions of Refs. \cite{Lahde:2002wj,Li:2007xr}, while the result for $\psi(2S)\rightarrow\gamma\eta_c(1S)$ shows good agreement with the literature \cite{Lahde:2002wj}. However, theoretical investigations into cases involving highly excited states are relatively scarce, and the results exhibit significant discrepancies. All of this indicates that the hindered decay processes of this type call for more theoretical and experimental attention.

\subsection{The electromagnetic multipole contributions of charmonium decays}
As mentioned earlier, our calculation of the electromagnetic transition amplitude includes multipole contributions from $M1$, $E2$, $M3$, and $E4$. Here, we examine in detail the contributions of each multipole transition. And discuss these separately for the $\psi\rightarrow\gamma\eta_c$ and $\eta_c\rightarrow\gamma\psi$ decays.
\subsubsection{$\psi\rightarrow\gamma\eta_c$}
In Table \ref{pai1s}, the first column presents the results for the total decay width (denoted by $EM$) of $\psi(nS, {}^3D_1) \to \gamma \eta_c(mS)$ ($n \geq m$), corresponding to the combined contribution of $M1 + E2 + M3 + E4$. The subsequent columns provide the individual contributions of $M1$, $E2$, $M3$, and $E4$ to the decay width, respectively. Due to space limitations, the table does not include contributions from interference terms between different transitions, such as $M1 \times E2$, etc., as the individual contributions of $M1$, $E2$, $M3$, and $E4$ presented are sufficient to illustrate the key points.

\begin{table}[h]
\begin{center}
\caption{Contributions from individual multipole transitions to the $\psi\to\gamma\eta_c$ decay width (keV), $EM$ corresponds to the total decay width.}\label{pai1s}
{\begin{tabular}{|c|c|c|c|c|c|} \hline\hline \diagbox {$1^{--}$}{$0^{-+}$}
                                            &~~~~EM ~~~~      & M1          &E2                &M3                       &E4                 \\ \hline
~~$\psi(1S)\rightarrow\gamma\eta_c(1S)$~    &1.46        &0.375       &0.276      &~~~$2.67\times10^{-3}$~~   &$3.00\times10^{-4}$      \\ \hline
$\psi(2S)\rightarrow\gamma\eta_c(1S)$       &1.20        &0.383       &1.37               &0.243                    &0.0386            \\ \hline
$\psi(2S)\rightarrow\gamma\eta_c(2S)$       &0.135       &0.0245      &0.0334        &$3.60\times10^{-4}$  &~~~$7.64\times10^{-5}$~~~   \\ \hline
$\psi(3770)\rightarrow\gamma\eta_c(1S)$     &1.54        &0.0360      &0.240               &0.758                     &$5.18\times10^{-3}$            \\ \hline
$\psi(3770)\rightarrow\gamma\eta_c(2S)$   &~~0.0233~~&~~$4.33\times10^{-7}$~~&~~$5.51\times10^{-3}$~~&~~$5.97\times10^{-3}$  &$2.72\times10^{-6}$      \\ \hline
$\psi(3S)\rightarrow\gamma\eta_c(1S)$       &0.0122  &~~$3.04\times10^{-3}$~~ &0.147  &0.0194          &0.0355            \\ \hline
$\psi(3S)\rightarrow\gamma\eta_c(2S)$       &0.140      &0.0325      &0.241              &0.0402              &$9.30\times10^{-3}$   \\ \hline
$\psi(3S)\rightarrow\gamma\eta_c(3S)$       &0.864      &0.145       &0.215              &$2.56\times10^{-3}$ &$1.18\times10^{-3}$   \\ \hline\hline
\end{tabular}}
\end{center}
\end{table}

As can be seen from Table \ref{pai1s}, the higher-order multipole contributions, i.e., the relativistic corrections, play a dominant role in all the decay processes $\psi(nS, {}^3D_1) \to \gamma \eta_c(mS)$. From the perspective of multipole contributions, in the process $\psi(1S)\rightarrow\gamma\eta_c(1S)$, the non-relativistic $M1$ contribution 0.375 keV is the largest, followed by the $E2$ contribution 0.276 keV, whereas the $M3$ and $E4$ contributions are negligible compared to the former two. Nevertheless, the non-relativistic width of 0.375 keV differs significantly from the final result of 1.46 keV, indicating substantial relativistic effects in this process. If we define the relativistic effect as $\frac{\Gamma_{rel}-\Gamma_{non-rel}}{\Gamma_{rel}}$, where $\Gamma_{rel}$ is our total decay width and $\Gamma_{non-rel}$ is the contribution of $M1$, then the relativistic effect for this process amounts to $74.3\%$.

Apart from $\psi(1S)\rightarrow\gamma\eta_c(1S)$, in all other $\psi(nS) \to \gamma \eta_c(mS)$ processes, the $E2$ contribution is the largest. For example, in the $\psi(2S)\rightarrow\gamma\eta_c(1S)$ decay, the $E2$ contribution is 1.47 keV, very close to the total result of 1.20 keV, followed by the $M1$ contribution of 0.383 keV. From Table \ref{incharm}, we note that $E2$ corresponds to $S\times P' + P\times S'$, whereas $M1$ corresponds to $S\times S'$. This indicates that the overlap integral between the lowest-order $S$-waves is suppressed and smaller than the overlap integral between the next-order $S\times P' + P\times S'$. It is pointed out in Ref. \cite{Eichten:2007qx} that the orthogonality between the $1S$ and $2S$ wave functions occurs in the limit of zero photon energy, which is consistent with our results.
The relativistic effects for the decay processes $\psi(2S)\rightarrow\gamma\eta_c(1S,~ 2S)$ are (68.1\%,~81.9\%), for $\psi(3S)\rightarrow\gamma\eta_c(1S,~ 2S,~3S)$ are (75.1\%,~76.8\%,~83.2\%).

In contrast, for the decay $\psi(3770) \to \gamma \eta_c(mS)$, the $M3$ contribution is the largest, followed by the $E2$ term, while that from $E4$ is very small and negligible. For instance, in the decay $\psi(3770) \to \gamma \eta_c(1S)$, the $M3$ contribution of 0.758 keV is greater than the $E2$ contribution of 0.240 keV, and significantly exceeds the $M1$ contribution of 0.0360 keV.
The theoretical decay width of $\psi(3770) \to \gamma \eta_c(2S)$ is significantly smaller than that of $\psi(3770) \to \gamma \eta_c(1S)$. Moreover, the same relationship holds for their respective multipole contributions. This suppression arises because the $\eta_c(2S)$ wave function possesses a node. When computing the overlap integral of the wave functions, the $\eta_c(2S)$ wave function on either side of the node provides contributions with opposite signs. These contributions cancel each other out, resulting in a very small width. Finally, the calculated relativistic effect for $\psi(3770) \to \gamma \eta_c(1S)$ is 97.7\%, while for $\psi(3770) \to \gamma \eta_c(2S)$ it is almost 100\%.

\subsubsection{$\eta_c\rightarrow\gamma\psi$}

\begin{table}[h]
\begin{center}
\caption{Contributions from individual multipole transitions to the $\eta_c$$\to$$\gamma\psi$  decay width (keV).}\label{etac}
{\begin{tabular}{|c|c|c|c|c|c|} \hline\hline \diagbox {$0^{-+}$}{$1^{--}$}
                  &~~EM decay~~    &~~M1 decay~~    &~~E2 decay~~    &M3 decay  &E4 decay       \\ \hline
~~~$\eta_c(2S)\rightarrow\gamma\psi(1S)$~~~    &0.630         &0.387        &0.954              &0.582      &$1.85\times10^{-3}$      \\ \hline
$\eta_c(3S)\rightarrow\gamma\psi(1S)$          &0.413         &0.0376       &0.326   &$1.05\times10^{-4}$     &0.0653      \\ \hline
$\eta_c(3S)\rightarrow\gamma\psi(2S)$          &0.112         &0.0301         &0.224    &$8.79\times10^{-6}$     &$1.46\times10^{-3}$      \\ \hline
$\eta_c(3S)\rightarrow\gamma\psi(1D)$          &0.00316 &$6.89\times10^{-6}$ &0.0146    &$6.80\times10^{-4}$     &0.0236           \\ \hline
\hline
\end{tabular}}
\end{center}
\end{table}

From Table \ref{etac}, the contribution of $E2$ transition in process $\eta_c(2S)\rightarrow\gamma\psi(1S)$ is relatively large, and the $E2$ transition is the dominant transition in the electromagnetic decay of $\eta_c(3S)$. Similar to $\psi\rightarrow\gamma\eta_c$, the higher multipole transition contributions of $\eta_c\rightarrow\gamma\psi$ still make significant contributions. And the relativistic effect of $\eta_c(2S)\rightarrow\gamma\psi(1S)$ and $\eta_c(3S)\rightarrow\gamma\psi(1S,2S)$ are 38.6\% and (90.9\%,~ 73.1\%), respectively.

In the multipole contributions to the transition $\eta_c(3S)\rightarrow\gamma\psi(1D)$, the $E4$ contribution dominates, followed by the $E2$ contribution, while the non-relativistic $M1$ contribution is minimal. This demonstrates that nearly all contributions originate from relativistic corrections.

\subsection{The EM decays of bottomonium}

The results of the electromagnetic radiative decays $\Upsilon\rightarrow\gamma\eta_b$ and $\eta_b\rightarrow\gamma\Upsilon$ are presented in Table \ref{bottom}, which also includes other theoretical predictions and experimental data from PDG for comparison. In theoretical studies, various models have been employed,
including the model of potential NRQCD \cite{Pineda:2013lta}, the relativistic quark model which the decay rates calculated discarding all relativistic corrections $\Gamma_{NR}$ and the relativistic quark model \cite{Ebert:2002pp}, the relativistic impulse approximation (RIA) and scalar confining components in the framework of the covariant Blankenbecler-Sugar (BSLT) equation \cite{Lahde:2002wj}, and the NRQCD model \cite{QuarkoniumWorkingGroup:2004kpm}.

\begin{table}[htp]
\begin{center}
\caption{The EM decay widths (eV) of bottomonium.} \label{bottom}
\begin{tabular}{|c|c|c|c|c|c|c|c|}
\hline\hline
\textbf{Process}&\textbf{~~~Ours~~~}&\textbf{~~~\cite{Pineda:2013lta}}~~~&\textbf{~~$\Gamma_{NR}$\cite{Ebert:2002pp}}~~&\textbf{~~~\cite{Ebert:2002pp}}~~~&\textbf{~~\cite{Lahde:2002wj}}~~ &\textbf{~~~\cite{QuarkoniumWorkingGroup:2004kpm}}~~~& \textbf{PDG \cite{PDG}}\\
\hline
$\Upsilon(1S)\rightarrow\gamma\eta_b(1S)$  &15.1     &15.18 &9.7   &5.8  &7.7           &8.953  &                   \\
\hline
$\Upsilon(2S)\rightarrow\gamma\eta_b(1S)$  &11.6     &6     &1.3   &6.4  &11.0          &2.809  &$17.6^{+5.0}_{-4.3}$ \\
$\Upsilon(2S)\rightarrow\gamma\eta_b(2S)$  &0.550    &0.668 &1.6   &1.4  &0.53          &1.509  &                   \\
\hline
$\Upsilon(1D)\rightarrow\gamma\eta_b(1S)$  &0.196    &      &      &     &              &       &                   \\
$\Upsilon(1D)\rightarrow\gamma\eta_b(2S)$  &0.0729   &      &      &     &              &       &                   \\
\hline
$\Upsilon(3S)\rightarrow\gamma\eta_b(1S)$  &10.8     &      &2.5   &10.5 &7.3           &2.435  &$10.4\pm 2.4$\\
$\Upsilon(3S)\rightarrow\gamma\eta_b(2S)$  &3.90     &      &0.2   &1.5  &2.9           &0.707  &$<12.6$            \\
$\Upsilon(3S)\rightarrow\gamma\eta_b(3S)$  &0.260    &      &0.9   &0.8  &0.13          &0.826  &                   \\
\hline
$\eta_b(2S)\rightarrow\gamma\Upsilon(1S)$  &8.06     &80    &2.4   &11.8 &102           &2.832  &                   \\
\hline
$\eta_b(3S)\rightarrow\gamma\Upsilon(1S)$  &12.9     &      &5.8   &24.0 &106           &3.757  &                   \\
$\eta_b(3S)\rightarrow\gamma\Upsilon(2S)$  &3.58     &      &0.4   &2.8  &12.6          &0.620  &                   \\
$\eta_b(3S)\rightarrow\gamma\Upsilon(1D)$  &0.685    &      &      &     &              &       &                   \\
\hline\hline
\end{tabular}
\end{center}
\end{table}

As can be seen from the Table \ref{bottom}, experimental studies thus far have been scarce, with only a few articles investigating $\Upsilon(2S)\rightarrow\gamma\eta_b(1S)$ \cite{BaBar:2009xir,Belle:2018aht}, $\Upsilon(3S)\rightarrow\gamma\eta_b(1S)$ \cite{CLEO:2009nxu,BaBar:2009xir,BaBar:2008dae}, and $\Upsilon(3S)\rightarrow\gamma\eta_b(2S)$ \cite{CLEO:2004jkt,BaBar:2011xka}, etc. Moreover, these studies exhibit large uncertainties, with the latter process only providing an upper limit. For the decay $\Upsilon(2S)\rightarrow\gamma\eta_b(1S)$, our theoretical result of 11.6 eV is close to the lower limit reported in the PDG. For the process $\Upsilon(3S)\rightarrow\gamma\eta_b(1S)$, our result 10.8 eV is in good agreement with the experimental data $10.4\pm 2.4$ eV.

Unlike radiative decay processes where the $E1$ transition serves as the leading-order term, such as $\chi_b \to \Upsilon$ or $\Upsilon \to \chi_b$, which exhibit small relativistic corrections, in these cases the non-relativistic $E1$ amplitude dominates, resulting in relatively consistent predictions across theoretical models  \cite{Pei:2024hzv}. In contrast, for the processes studied in this paper, $\eta_b \to \Upsilon$ and $\Upsilon \to \eta_b$, see Table \ref{bottom}, theoretical results show significant discrepancies. This, to some extent, reflects the substantial relativistic corrections required for these processes.
This is also evident from the large difference between the non-relativistic result and the relativistic result for $\eta_b \to \Upsilon$ and $\Upsilon \to \eta_b$ given in Ref. \cite{Ebert:2002pp} within Table \ref{bottom}. Furthermore, studies in Ref. \cite{Godfrey:2001eb} also indicated that relativistic corrections play an important role in processes $\Upsilon(2S)\rightarrow\gamma\eta_b(1S)$ and $\Upsilon(3S)\rightarrow\gamma\eta_b(1S,2S)$.

\subsection{The electromagnetic multipole contributions of bottomonium decays}

Large relativistic corrections imply significant contributions from higher multipoles in electromagnetic radiative transition processes. This is because our transition amplitude is $M1 + E2 + M3 + E4$, where the $M1$ transition is non-relativistic, while $E2$, $M3$, and $E4$ are relativistic. Subsequently, we will present the individual contributions of each multipole transition ($M1$, $E2$, $M3$, $E4$) for the decays $\eta_b \to \Upsilon$ and $\Upsilon \to \eta_b$, respectively. Additionally, the relativistic effects of these decay processes will be discussed.

\subsubsection{$\Upsilon\rightarrow\gamma\eta_b$}

From the comparison of the total EM results and the non-relativistic $M1$ contributions in Table \ref{bottom1S}, it is evident that relativistic corrections still play the dominant role in the radiative decays of bottomonium. However, unlike the charmonium case, in the decays $\Upsilon(nS)\rightarrow\gamma\eta_b(mS)$ ($n\geq m$), only for $\Upsilon(2S)\rightarrow\gamma\eta_b(1S)$ and $\Upsilon(3S)\rightarrow\gamma\eta_b(1S)$ does the $E2$ contribution dominate, followed by the non-relativistic $M1$. In all other processes, the $M1$ gives the largest contribution, followed by $E2$. In contrast, for charmonium, only in $\psi(1S)\rightarrow\gamma\eta_c(1S)$ does $M1$ give the largest contribution; in all other cases, $E2$ dominates. This phenomenon is consistent with the bottom quark being heavier than the charm quark. The relativistic effects are 72.8\% for $\Upsilon(1S) \rightarrow \gamma\eta_b(1S)$, (65.9\%, 74.4\%) for $\Upsilon(2S) \rightarrow \gamma\eta_b(1S,2S)$, and (68.1\%, 70\%, 75.2\%) for $\Upsilon(3S) \rightarrow \gamma\eta_b(1S,2S,3S)$. This demonstrates that although bottomonium is very heavy and its relativistic corrections are generally considered small and negligible in ordinary processes, they nevertheless turn out to be dominant in these specific decay channels.

\begin{table}[H]
\begin{center}
\caption{Contributions from individual multipole transitions to the  $\Upsilon$$\to$$\gamma\eta_b$ decay width (eV).}\label{bottom1S}
{\begin{tabular}{|c|c|c|c|c|c|} \hline\hline \diagbox {$1^{--}$}{$0^{-+}$}
                                                 &~~EM decay~~    &~~~~~M1 decay~~~~~         &E2 decay~              &M3 decay                   &E4 decay             \\ \hline
~~$\Upsilon(1S)\rightarrow\gamma\eta_b(1S)$~~      &15.1          &4.11             &3.24  &~~$1.65\times10^{-3}$~~ &$2.11\times10^{-4}$  \\ \hline
$\Upsilon(2S)\rightarrow\gamma\eta_b(1S)$          &11.6          &3.96             &6.07            &0.478                 &0.124          \\ \hline
$\Upsilon(2S)\rightarrow\gamma\eta_b(2S)$          &0.550         &0.141            &0.123    &$7.62\times10^{-8}$&~~$3.31\times10^{-8}$~~\\ \hline
$\Upsilon(1D)\rightarrow\gamma\eta_b(1S)$          &0.196         &0.437            &0.0518          &0.819          &$8.01\times10^{-4}$   \\ \hline
$\Upsilon(1D)\rightarrow\gamma\eta_b(2S)$&~~0.0729~~&~~$2.53\times10^{-5}$~~&$~~3.71\times10^{-4}$~~&0.0224            &0.0181        \\ \hline
$\Upsilon(3S)\rightarrow\gamma\eta_b(1S)$          &10.8          &3.44             &5.43            &0.220                  &0.189         \\ \hline
$\Upsilon(3S)\rightarrow\gamma\eta_b(2S)$          &3.90          &1.17             &0.415           &$0.115\times10^{-3}$       &0.00475        \\ \hline
$\Upsilon(3S)\rightarrow\gamma\eta_b(3S)$          &0.260         &0.0644           &0.0586   &$4.96\times10^{-5}$  &$4.37\times10^{-5}$ \\ \hline\hline
\end{tabular}}
\end{center}
\end{table}

Analogous to the case of $\psi(1D)$, the M3 component dominates in the transition $\Upsilon(1D)\rightarrow\gamma\eta_b(1S,2S)$. Furthermore, due to the nodal structure in the final-state wave function, both the total width of $\Upsilon(1D)\rightarrow\gamma\eta_b(2S)$ and the contributions from individual multipole transitions are significantly smaller than those in the nodeless $\Upsilon(1D)\rightarrow\gamma\eta_b(1S)$ process.

\subsubsection{$\eta_b\rightarrow\gamma\Upsilon$}

Table \ref{bottom0S} presents the total decay widths and the contributions from individual multipole transitions for $\eta_b \rightarrow \gamma\Upsilon$. For the decays $\eta_b(nS)\rightarrow\gamma\Upsilon(mS)$ ($n>m$), the M1 contribution plays the dominant role and is significantly larger than those from other multipoles. This indicates that the non-relativistic contributions constitute the primary component in these processes. This differs significantly from the corresponding charmonium process where the $E2$ contribution dominates, reflecting the kinematic difference attributed to the bottom quark being heavier than the charm quark. The relativistic effects are 38.3\% for $\eta_b(2S)\rightarrow\gamma\Upsilon(1S)$ and (45.0\%, 51.7\%) for $\eta_b(3S)\rightarrow\gamma\Upsilon(1S,2S)$.

In the decay $\eta_b(3S)\rightarrow\gamma\Upsilon(1D)$, the $E2$ transition provides the largest contribution, followed by $M3$, while the $M1$ contribution is the smallest, even smaller than that of $E4$. This indicates that the non-relativistic contribution to this process is very small, and nearly all of the contribution originates from relativistic corrections.

\begin{table}[H]
\begin{center}
\caption{Contributions from individual multipole transitions to the $\eta_b$$\to$$\gamma\Upsilon$ decay width (eV).}\label{bottom0S}
{\begin{tabular}{|c|c|c|c|c|c|} \hline\hline \diagbox {$0^{-+}$}{$1^{--}$}
                                             &~~~EM decay       &~~~M1 decay~~~  &~~~E2 decay~~~          &M3 decay                    &E4 decay            \\ \hline
~~~~$\eta_b(2S)\rightarrow\gamma\Upsilon(1S)$~~~~&8.06        &~~~4.97~~~     &0.137   &~~~$1.19\times10^{-4}$~~~       &0.0515          \\ \hline
$\eta_b(3S)\rightarrow\gamma\Upsilon(1S)$          &12.9            &7.09     &~~~0.968~~~  &$8.23\times10^{-5}$    &$3.94\times10^{-3}$  \\ \hline
$\eta_b(3S)\rightarrow\gamma\Upsilon(2S)$          &3.58            &1.73         &0.312      &$7.88\times10^{-6}$  &~~$3.09\times10^{-4}$~~\\ \hline
$\eta_b(3S)\rightarrow\gamma\Upsilon(1D)$      &~~~0.685~~~&$7.71\times10^{-4}$ &0.365            &0.153                 &0.0384      \\ \hline\hline
\end{tabular}}
\end{center}
\end{table}

{
\subsection{Discussions}

The transition amplitude can also be expressed in terms of helicity amplitudes \cite{prd,prl}.
For processes such as $\psi(nS) \rightarrow\eta_{c}(mS)+\gamma$ studied in this paper, parity invariance restricts the number of independent helicity amplitudes to one.
It is generally believed that the number of independent multipole amplitudes must equal the number of independent helicity amplitudes because of rotational invariance. In this case, it appears that only a single $M1$ multipole amplitude should contribute, rather than four in this paper. The cause of this confusion lies in the assumption that orbital angular momentum is a good quantum number, in which case there exists a one-to-one correspondence between total angular momentum and orbital angular momentum.

Due to rotational invariance and angular momentum conservation, Ref. \cite{prd} provides the conversion relationship between the helicity amplitude and the multipole amplitude:
\begin{eqnarray}
A^{(J)}_{|\nu|}=\sum\limits_{J_{\gamma}}a^{(J)}_{J_{\gamma}}(\frac{2J_{\gamma}+1}{2J+1})^{1/2}\left(J_{\gamma},1,1,|\nu|-1\Big|J,|\nu|\right),
%\nonumber\\
%B^{(J)}_{|\nu'|}=\sum\limits_{J'_{\gamma}}b^{(J)}_{J'_{\gamma}}(\frac{2J'_{\gamma}+1}{2J+1})^{1/2}\left(J'_{\gamma},1,1,|\nu'|-1,\Big|J,|\nu'|\right)
\end{eqnarray}
where $A^{(J)}_{|\nu|}$ represents the helicity amplitude, while $a^{(J)}_{J_{\gamma}}$  is related to the ``multipole amplitude". This formula shows that the number of helicity amplitudes is equal to the number of allowed total angular momenta $J_{\gamma}$ (and does not directly correspond to the number of multipole amplitudes). If there is a one-to-one correspondence between $J_{\gamma}$ and $L$, that is $a^{(J)}_{J_{\gamma}}=a^{(J)}_{J_{\gamma}=L+n}$ ($n=0, \pm 1$), then the number of helicity amplitudes and multipole amplitudes is equal, and in this case, $a^{(J)}_{J_{\gamma}}$  exactly denotes the multipole amplitude.

However, since orbital angular momentum is generally not a good quantum number, there is typically a one-to-many correspondence between total angular momentum and orbital angular momentum especially in relativistic transition amplitudes. So the number of $L$ is greater than the corresponding $J_{\gamma}$, and more number of $L$ leads to more number of multipole amplitudes (there is a direct one-to-one correspondence between them), therefore the number of multipole amplitudes is generally larger than the number of helicity amplitudes. In this case, $a^{(J)}_{J_{\gamma}}$ should be written as $a^{(J)}_{J_{\gamma},L_1}+a^{(J)}_{J_{\gamma},L_2}+...$. For the electromagnetic transitions between $0^{-+}$ and $1^{--}$ mesons, we obtain four multipole amplitudes, $M1$, $E2$, $M3$, and $E4$, rather than only the $M1$ transition.

We note that in Ref. \cite{close}, alongside the electric dipole transition $E1$, there exists a term referred to as the ``extra" electric dipole $E_R$ (e.g., in the $^3S_1\to ^3P_0$ transition which has only one helicity amplitude). The authors stated that, unlike the $E1$, the $E_R$ is ``spin-flip" in nature, just like a magnetic transition. Ref. \cite{Deng:2015bva} pointed out that this $E_R$ term belongs to magnetic transition. This confusion can be well explained by the results of this paper, which show that the $^3S_1\to {}^3P_0$ transition includes not only $E1$ but also contributions from $M2$, among others.
}

\section{Summary}
In this paper, we employ the BS equation approach to study the electromagnetic decays of heavy quarkonia. Unlike the typical non-relativistic approach which only calculates the $M1$ transition, we present the transition amplitude as $M1+E2+M3+E4$, where $M1$ is non-relativistic while the other multipole transitions are relativistic.

Our results demonstrate that: First, the processes investigated in this work, whether charmonium or bottomonium decays, exhibit pronounced relativistic effects, marking a distinct departure from conventional processes.

Second, among the $\eta_b(nS)\rightarrow\gamma\Upsilon(mS)$ ($n>m$) and $\Upsilon(n'S)\rightarrow\gamma\eta_b(m'S)$ ($n'\geq m'$) processes, only $\Upsilon(2S)\rightarrow\gamma\eta_b(1S)$ and $\Upsilon(3S)\rightarrow\gamma\eta_b(1S)$ exhibit $E2$ dominance, while all others are dominated by $M1$ contributions. In contrast, for the decays $\eta_c(nS)\rightarrow\gamma\psi(mS)$ and $\psi(n'S)\rightarrow\gamma\eta_c(m'S)$, only $\psi(1S)\rightarrow\gamma\eta_c(1S)$ is dominated by the $M1$ transition; all others are dominated by $E2$. This difference reflects a kinematic distinction, namely that the bottom quark is heavier than the charm quark.

Finally, when examining relativistic effects, most processes show no significant reflection of the mass difference between charmonia and bottomonia.
For instance, the relativistic correction in $\eta_b(2S) \rightarrow \gamma\Upsilon(1S)$ amounts to $38.3\%$, closely matching the $38.6\%$ obtained in $\eta_c(2S)\rightarrow \gamma\psi(1S)$. Meanwhile, decays $\Upsilon(nS)\rightarrow\gamma\eta_b(mS)$ ($n\geq m$) exhibit relativistic effects ranging from $65.9\%$ to $75.2\%$, strikingly similar to the $68.1\%\sim83.2\%$ range found in analogous charmonium processes $\psi(nS)\rightarrow\gamma\eta_c(mS)$.
These observations indicate that for these processes, the magnitude of the relativistic effects primarily depends on dynamical effects and is essentially irrelevant to kinematical effects related to quark mass.

{\bf Acknowledgments}
This work is supported by the National Natural Science Foundation of China (NSFC) under the Grants No. 12575097 and No. 12075073. W. Li is also supported by Hebei Agricultural University introduced talent research special project (No. YJ2024038). Q. Li is supported by the National Key R\&D Program of China\,(2022YFA1604803) and the Natural Science Basic Research Program of Shaanxi\,(No.\,2025JC-YBMS-020).

%\bibliography{references}{}

 \end{document}